\documentclass[letterpaper,aps,prc,superscriptaddress,showpacs,nofootinbib,floatfix,twocolumn]{revtex4-1}

\usepackage{hyperref}
\usepackage{color}
\usepackage{graphicx}	
\graphicspath{%
  {./},%
}

\usepackage{xspace}	

\newcommand{\he}{\mbox{$^{3}{\rm He}$}\xspace}

\newcommand{\pA}{\mbox{p+${\rm Au}$}\xspace}
\newcommand{\pPb}{\mbox{p+${\rm Pb}$}\xspace}
\newcommand{\dA}{\mbox{d+${\rm Au}$}\xspace}
\newcommand{\tA}{\mbox{t+${\rm Au}$}\xspace}
\newcommand{\heA}{\mbox{$^{3}{\rm He}$+${\rm Au}$}\xspace}
\newcommand{\hePb}{\mbox{$^{3}{\rm He}$+${\rm Pb}$}\xspace}

\newcommand{\sqsn}{\mbox{$\sqrt{s_{_{NN}}}$}\xspace}

\begin{document}

\title{Exploiting Intrinsic Triangular Geometry in Relativistic \heA Collisions to Disentangle Medium Properties}

\author{J. L. Nagle, A. Adare, S. Beckman, T. Koblesky, J. Orjuela Koop, D. McGlinchey, P. Romatschke} \affiliation{University of Colorado at Boulder}
\email{jamie.nagle@colorado.edu}

\author{J. Carlson, J. E. Lynn, M. McCumber} \affiliation{Los Alamos National Laboratory}

\date{\today}

\begin{abstract}

Recent results in \dA and \pPb collisions at RHIC and the LHC provide evidence for collective
expansion and flow of the created medium.  We propose a control set of experiments to directly compare 
particle emission patterns from \pA, \dA, and \heA or \tA collisions at the same \sqsn.
Using Monte Carlo Glauber we find that a \he or triton projectile, with a realistic wavefunction description, 
induces a significant intrinsic triangular shape to the initial medium. 
If the system lives long enough, this survives into a significant third order flow moment $v_{3}$ even with viscous damping.
By comparing systems with one, two, and three initial hot spots,
one could disentangle the effects from the initial spatial distribution of the deposited energy and viscous damping.  
These are key tools to answering the question of how small a droplet of matter is necessary to form a quark-gluon 
plasma described by nearly inviscid hydrodynamics.

\end{abstract}

\maketitle

Nearly inviscid hydrodynamic expansion of a quark-gluon plasma followed by hadronization has become the standard
model for relativistic collisions of heavy nuclei at the Relativistic Heavy Ion Collider (RHIC) and the 
Large Hadron Collider (LHC)~\cite{Romatschke:2009im,Heinz:2013wva}.  
Fluctuations in the nucleon positions within the incident nuclei result
in an inhomogeneous distribution of initially deposited energy, and the influence of these spatial anisotropies survives into final state
hadron momentum distributions~\cite{Alver:2010gr}.  Measurements of such flow moments, $v_{2}$ for elliptic flow and $v_{3}$ 
for triangular flow for example, probe both the initial anisotropies and the viscous damping effect through the time
evolution of the medium.  Striking agreement between experimental data for higher order flow moments and viscous hydrodynamic 
calculations with lumpy initial conditions confirm values of the shear viscosity to entropy density $\eta/s = 1-2/4\pi$
~\cite{Schenke:2010rr,Schenke:2011bn}. 
Similar values for $\eta/s$ are also found in ultracold quantum gases and black holes, suggesting a much deeper connection of these strongly coupled systems \cite{Schafer:2009dj}.

Recent experimental results from central \dA and \pPb collisions at RHIC and the LHC, respectively, reveal remarkably similar ``flow'' 
patterns~\cite{Adare:2013piz,CMS:2012qk,Abelev:2012ola,Aad:2012gla},
contrary to expectations of forming no quark-gluon plasma from these small system collisions.
Qualitative agreement with the $v_{2}$ and $v_{3}$ results is obtained with 
hydrodynamics~\cite{Bozek:2011if,Qin:2013bha,Bzdak:2013zma}, 
though alternative explanations involving glasma diagrams~\cite{Dusling:2012iga} and other dynamics have also been proposed.  
The difference in both projectile (deuteron versus proton) and center-of-mass energy 
(\sqsn = 200 GeV versus \sqsn = 5.02 TeV) between the RHIC and LHC results provides an excellent lever arm for 
discriminating between underlying physics explanations, though ambiguities remain.

In this paper, we propose a set of control experiments that involve collisions of \pA, \dA, and \heA or \tA at the 
same \sqsn.  Such a set of experiments are available to run at RHIC with modest run lengths.  
We utilize detailed calculations of the \he and triton wavefunction for the initial distribution of nucleons
within the nuclei.  We then couple these distributions with Monte Carlo Glauber simulations to determine event-by-event
distributions of the deposited energy.  Individual events are then run through a modified version of the
relativistic viscous hydrodynamic transport code~\cite{Luzum:2008cw}, followed by a hadronization prescription and a hadron scattering transport code \cite{Novak:2013bqa}.  Final
distributions of $v_{2}$ and $v_{3}$ flow coefficients as a function of transverse momentum are calculated and compared 
between the colliding systems and with different input parameters, including $\eta/s$.  

As input to the Monte Carlo Glauber calculation~\cite{Miller:2007ri}, we require a realistic distribution of the nucleons
within the nuclei of interest.  For the $\rm{Au}$ nucleus, the nucleons are distributed following a standard Woods-Saxon 
distribution with radius and skin thickness parameters 6.42 fm and 
0.44 fm \cite{Hirano:2009ah}.
A hard-core repulsive potential is implemented as an exclusion radius of 0.4 fm between nucleons.  For the \dA
collision case, the deuteron is modeled via the Hulthen wavefunction 
(cf. Ref.~\cite{Adare:2013nff}).
In the case of \he and triton projectiles, the three-body dynamics are important to capture as we need to model the
distribution of the three hot spots created in collisions with $\rm{Au}$ nuclei.  
The \he and triton samples come from Green's function Monte Carlo calculations using the AV18 + UIX model interaction \cite{Carlson:1997qn}. These calculations correctly reproduce the measured charge radii
and form factors of these nuclei.
The relative distribution of proton pairs in \he also reproduces measurements of inclusive longitudinal
electron scattering. In practice, we use a database of 10,000 \he configurations which correctly sample the position of the three nucleons, including correlations.

We model collisions at \sqsn = 200 GeV with a nucleon-nucleon inelastic cross section of 42 mb and collisions at  \sqsn = 5.02 TeV with a nucleon-nucleon inelastic cross section of 70 mb.
For each individual event, to map the positions of the participating nucleons (those with at least one inelastic collision
in the event) to a distribution of energy deposited in the transverse plane, we assume that each participant contributes
an equal energy with a distribution that is Gaussian around its center point with $\sigma = 0.4$ fm, to match the
RMS radius of the nucleon.  There is an overall scale factor to convert these distributions to energy density, and this is determined by requiring our model to give multiplicities consistent with 
data in 0-5\% \dA collisions at \sqsn = 200 GeV and 0-5\% \pPb collisions at \sqsn = 5.02 TeV \cite{ABELEV:2013wsa}, respectively. For fixed \sqsn, the same factor is used for our \heA, \dA and \pA simulations because these systems are comparable in size and we have checked that viscous heating only changes the multiplicity by less than 11 percent for $\eta/s<2/4\pi$.

\begin{figure}[t]
  \centering
  \includegraphics[width=0.95\linewidth]{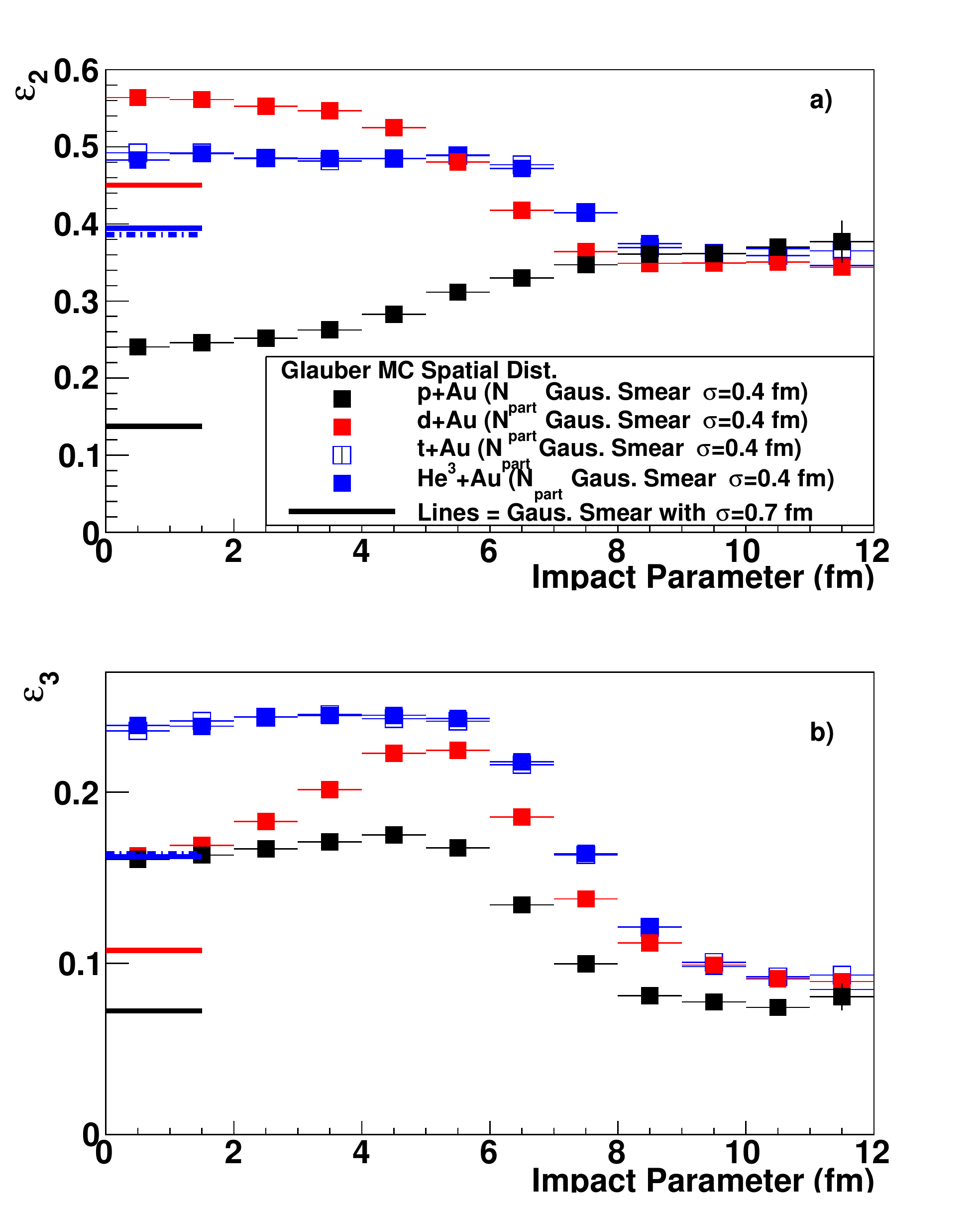}
  \caption{\label{fig:moments} (Color online) Monte Carlo Glauber results for the spatial
anisotropies $\varepsilon_{2}$ (panel a) and $\varepsilon_{3}$ (panel b) in \pA, \dA, \heA, and \tA collisions at $\sqrt{s_{NN}}=200$ GeV as a function
of impact parameter.  The points are calculated with a Gaussian smearing with $\sigma = 0.4$ fm for the energy distribution
from each participating nucleon. The lines are the results for central events with a larger Gaussian smearing
with $\sigma = 0.7$ fm.  
}\end{figure}

We have generated a million collision geometries for each case, \pA, \dA, \heA, and \tA,
and calculated the spatial anisotropy of the initial energy distribution using the following equation~\cite{Alver:2010gr}:
\begin{equation}
  \varepsilon_{n} = \frac{\sqrt{\left< r^{2}\cos(n\phi_{part})\right>^{2} + \left< r^{2} \sin(n\phi_{part})\right>^{2}}}{\left< r^{2} \right>}
\end{equation}
where $n$ is the $n$th moment of the spatial anisotropy calculated relative to the mean position. 
These distributions are calculated with respect to the axis associated with the $n$th moment as defined by:
\begin{equation}
  \psi_{n} = \frac{{\rm arctan}(\left< r^{2}\sin(n\phi_{part})\right>, \left< r^{2} \cos(n\phi_{part}) \right> ) + \pi}{n}.
\end{equation}

Figure~\ref{fig:moments} shows the $\varepsilon_{2}$ (elliptical) and $\varepsilon_{3}$ (triangular) event-averaged values
 as a function of impact parameter.  
For central events, small impact parameter, the $\varepsilon_{2}$ values are significantly larger for \dA as compared
with \pA since the deuteron typically creates two hot spots in the interaction creating a dumbbell shaped
energy distribution.   
The initial triangularity $\varepsilon_{3}$  is
largest in the \heA and \tA system.  It is notable
that the \pA and \dA central collisions induce the same $\varepsilon_{3}$ since they result only from fluctuations 
in contrast to the intrinsic triangularity in the \heA and \tA cases.  In all cases the differences between \tA and \heA
are negligible and we will only refer to \heA for the remainder of the paper.  
The lines indicate the change in the spatial anisotropies if the Gaussian smearing for each participant is 
increased to $\sigma = 0.7$ fm.  This has the largest impact on the \pA case as expected since it has the smallest
initial spatial scale.

We then run individual event initial conditions starting at a time $\tau=0.5$ fm/c through the well-tested boost-invariant relativistic viscous hydrodynamic evolution vh2~\cite{Luzum:2008cw}, modified by smearing the local energy density if it drops below one percent of the 
maximum value encountered at any particular instant in time. This smearing effectively avoids instabilities generated by the 
strong gradients present when simulating small non-homogeneous systems, while at the same time affecting bulk observables only on the per-mille level.

\begin{figure*}[t]
  \centering
  \includegraphics[width=0.95\textwidth]{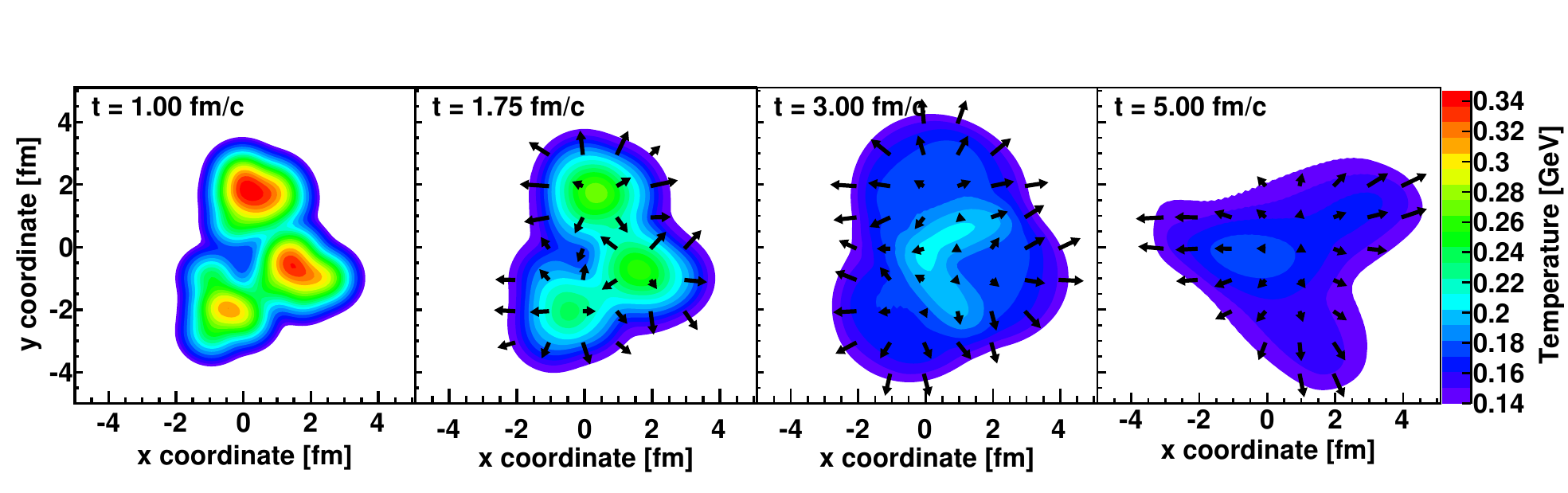}
  \caption{\label{fig:display} (Color online) An example time evolution of a \heA event from the initial
state to final state.  The color scale indicates the local temperature and the arrows are proportional
to the velocity of the fluid cell from which the arrow originates. 
}\end{figure*}

\begin{figure}[h]
  \centering
  \includegraphics[width=0.9\linewidth]{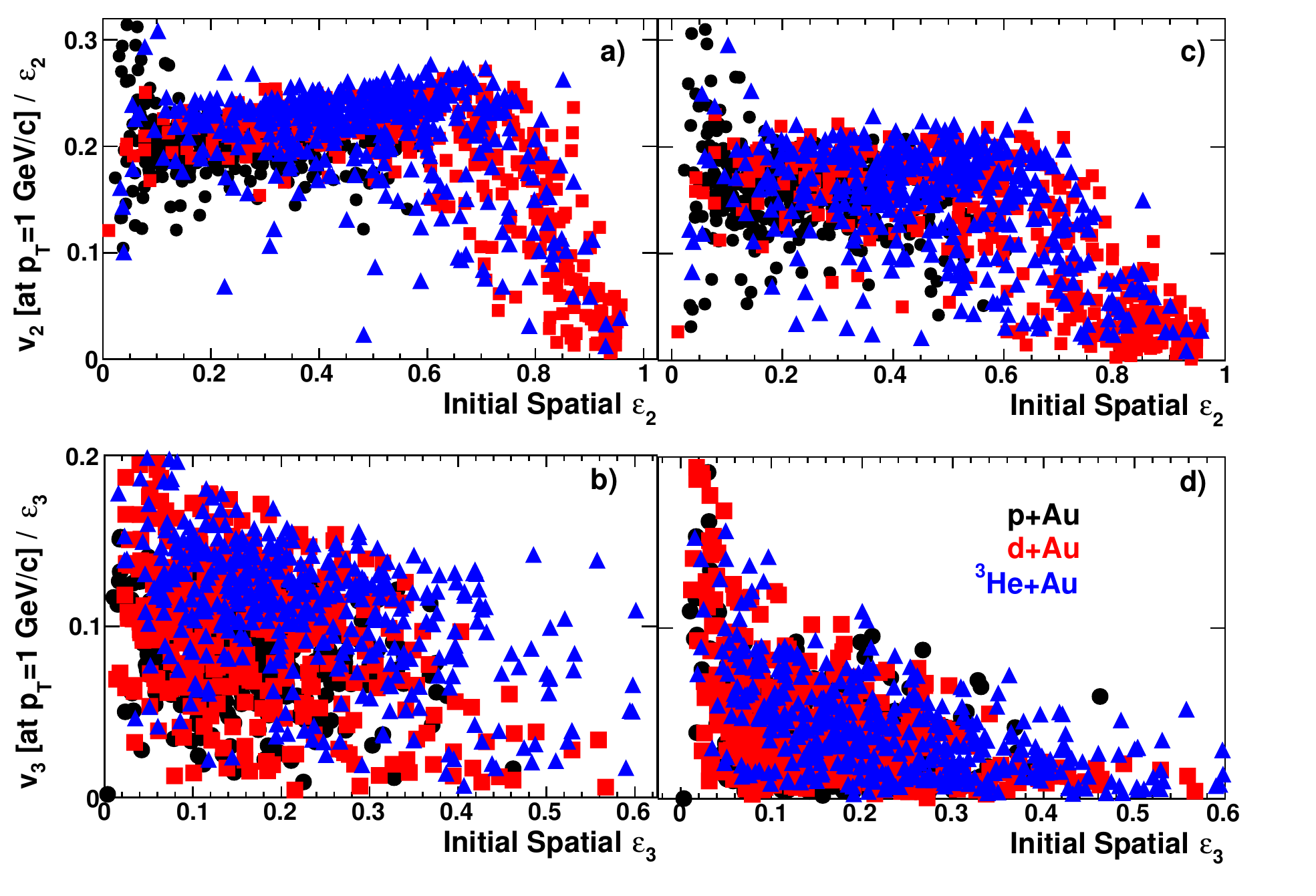}
  \caption{\label{fig:scaling} (Color online) $v_{n} / \varepsilon_{n}$ versus $\varepsilon_{n}$ with the flow
coefficient for pions evaluated at $p_{T} = 1.0$ GeV/c from \pA, \dA, and \heA central ($b < 2$ fm) events.
The results are with input parameters $\eta/s = 1/4\pi$ and initial Gaussian smearing $\sigma = 0.4$ fm and freeze-out temperatures of $T_F=150$ MeV (left) and $T_F=170$ MeV (right), respectively.
}\end{figure}

\begin{figure}[h]
  \centering
  \includegraphics[width=0.95\linewidth]{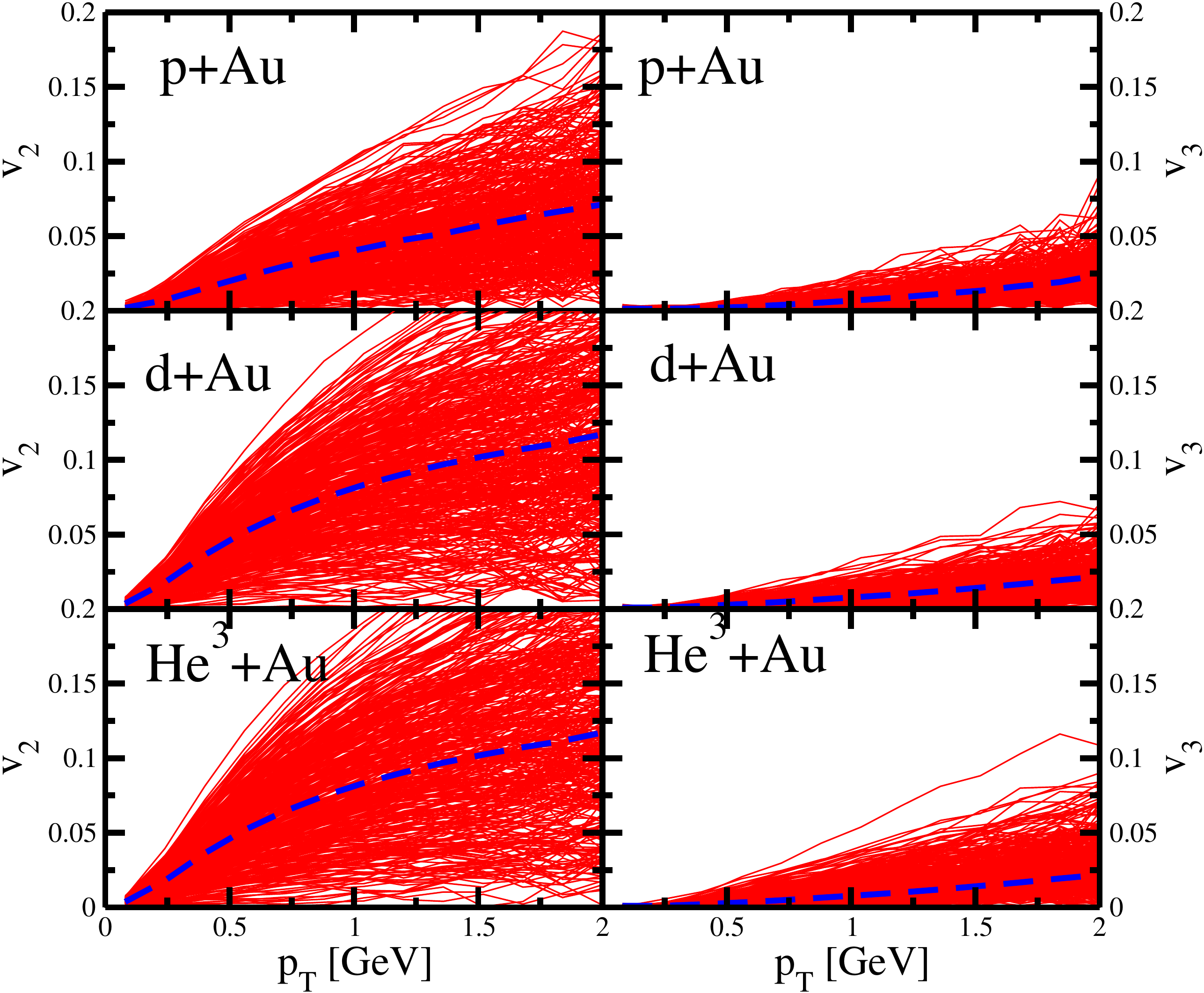}
  \caption{\label{fig:v23results} (Color online) Pion momentum
anisotropies $v_{2}$ and $v_{3}$ as a function of transverse momentum from individual \pA, \dA, and \heA central ($b < 2$ fm) events (full lines)
Dashed lines are the event-averaged values.
}\end{figure}

The results from an example \heA event are shown in Figure~\ref{fig:display}.  The first panel shows the
temperature profile, converted from energy density using a realistic QCD equation of state~\cite{Luzum:2008cw}, as generated
from the above described Monte Carlo Glauber.  This event has all three nucleons from the \he nucleus striking the
$\rm{Au}$ nucleus, thereby creating three hot spots.   In this event, the triangular initial spatial distribution
transforms into an inverted triangular distribution with maximal fluid velocity fields along the long axes of the
final triangular shape.

We have run thousands of individual events for \pA, \dA, and \heA with different values for the shear viscosity
and the initial spatial distribution smearing.  The final freeze-out hyper-surface of each event is then translated
into a distribution of hadrons via the Cooper-Frye freeze-out prescription~\cite{Cooper:1974mv}. In Figure~\ref{fig:scaling},
we compare the flow coefficients from the different systems and the scaling between initial spatial $\varepsilon_{n}$ moments and final state momentum $v_{n}$ values. Figure~\ref{fig:scaling}  shows
the pion $v_{n}$ at $p_{T} = 1.0$ GeV/c divided by $\varepsilon_{n}$ as a function of $\varepsilon_{n}$ for each individual
\pA, \dA, and \heA event,
for different freezeout temperatures $T_F$ controlling the lifetime of the system in the plasma phase.  The upper panels for $n=2$ shows a reasonably common scaling of $v_{2}/\varepsilon_{2}$ 
for all three systems with the \dA and \heA simply extending to larger eccentricities with only a modest dependence on $T_F$. 
There are a small set of events with very large $\varepsilon_{2}$, but then rather small final $v_{2}$.  Examination of these events reveals them to be \dA events where the two hot spots are so far apart that the hydrodynamic fluids never connect during the time evolution, and thus there is almost no elliptic flow.  There are a few \heA in this category where two nucleons are very close and the
third is quite far away, again having the same effect.

The lower panels for the $n=3$ case have lower values for $v_{3}/\varepsilon_{3}$ compared to $v_{2}/\varepsilon_{2}$ as
expected from larger viscous damping of higher moments.  There is significantly more spread of the individual events,
though an overall scaling is still observed.  
Even more dramatic is the dependence of $v_{3}/\varepsilon_{3}$ on $T_F$. Increasing $T_F$ from $150$ MeV to $170$ MeV considerably shortens the hydrodynamic evolution time and results in a strong reduction of $v_{3}/\varepsilon_{3}$ for all systems.

\begin{figure*}[t]
  \centering
  \includegraphics[width=0.32\textwidth]{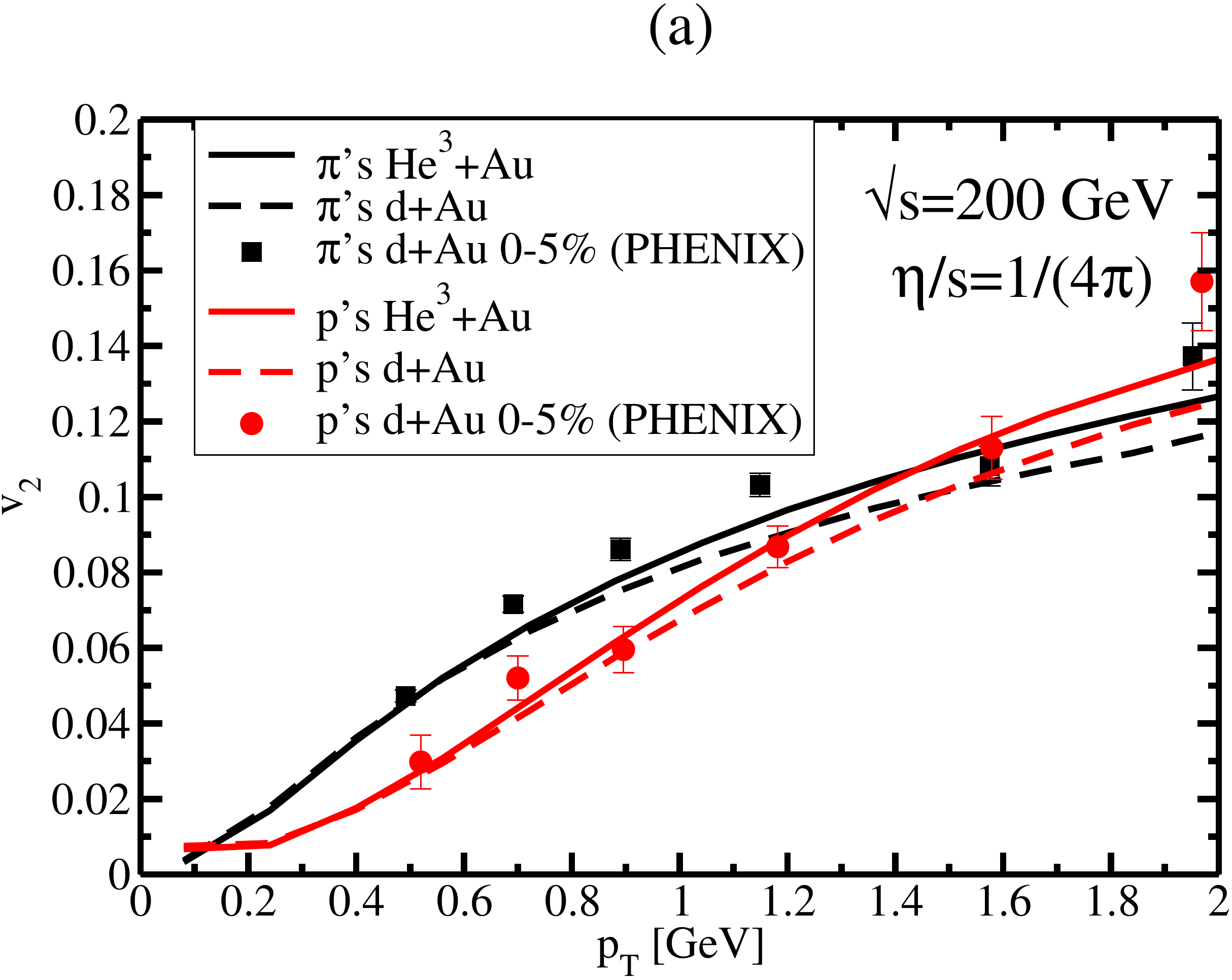}
  \includegraphics[width=0.32\textwidth]{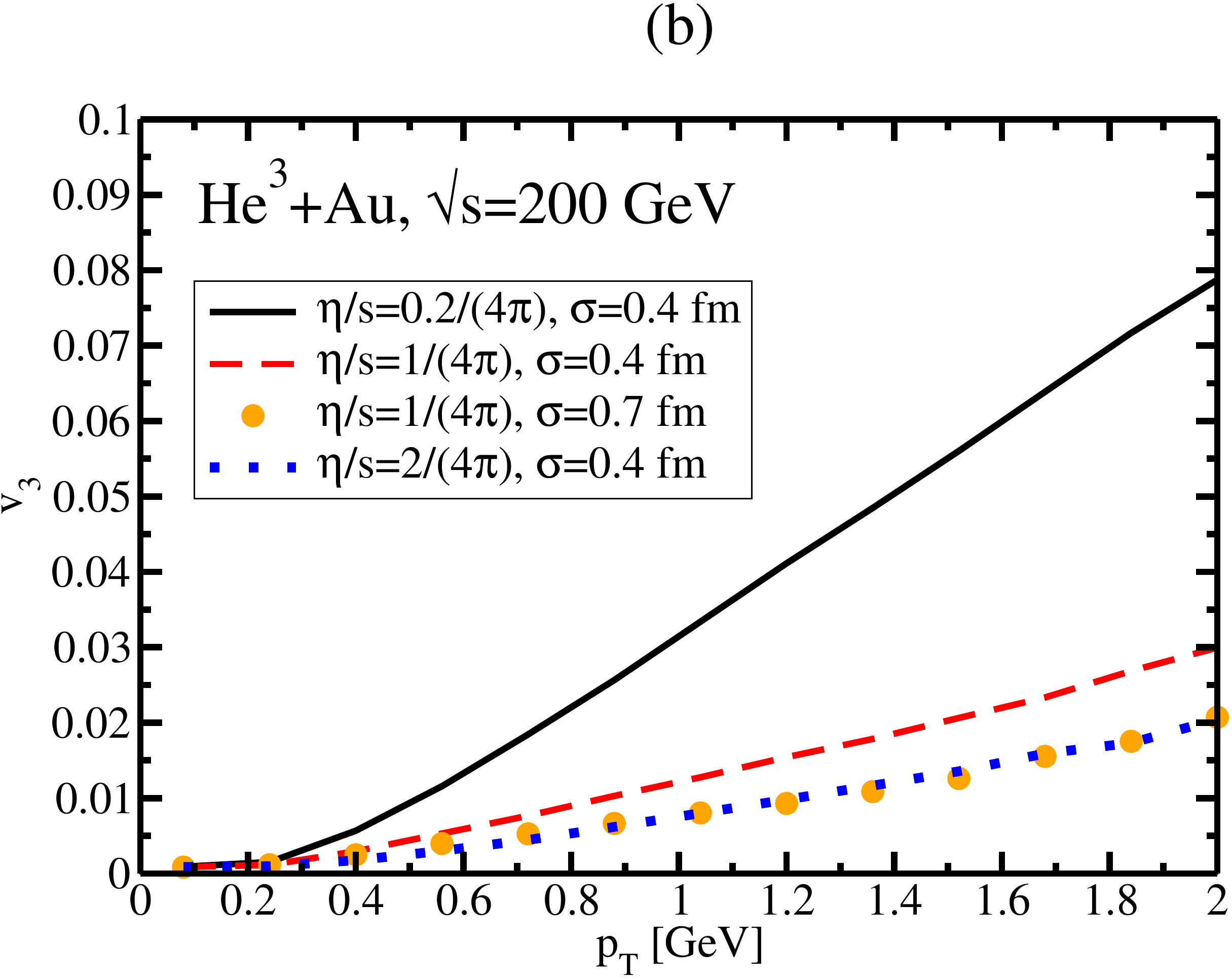}
  \includegraphics[width=0.32\textwidth]{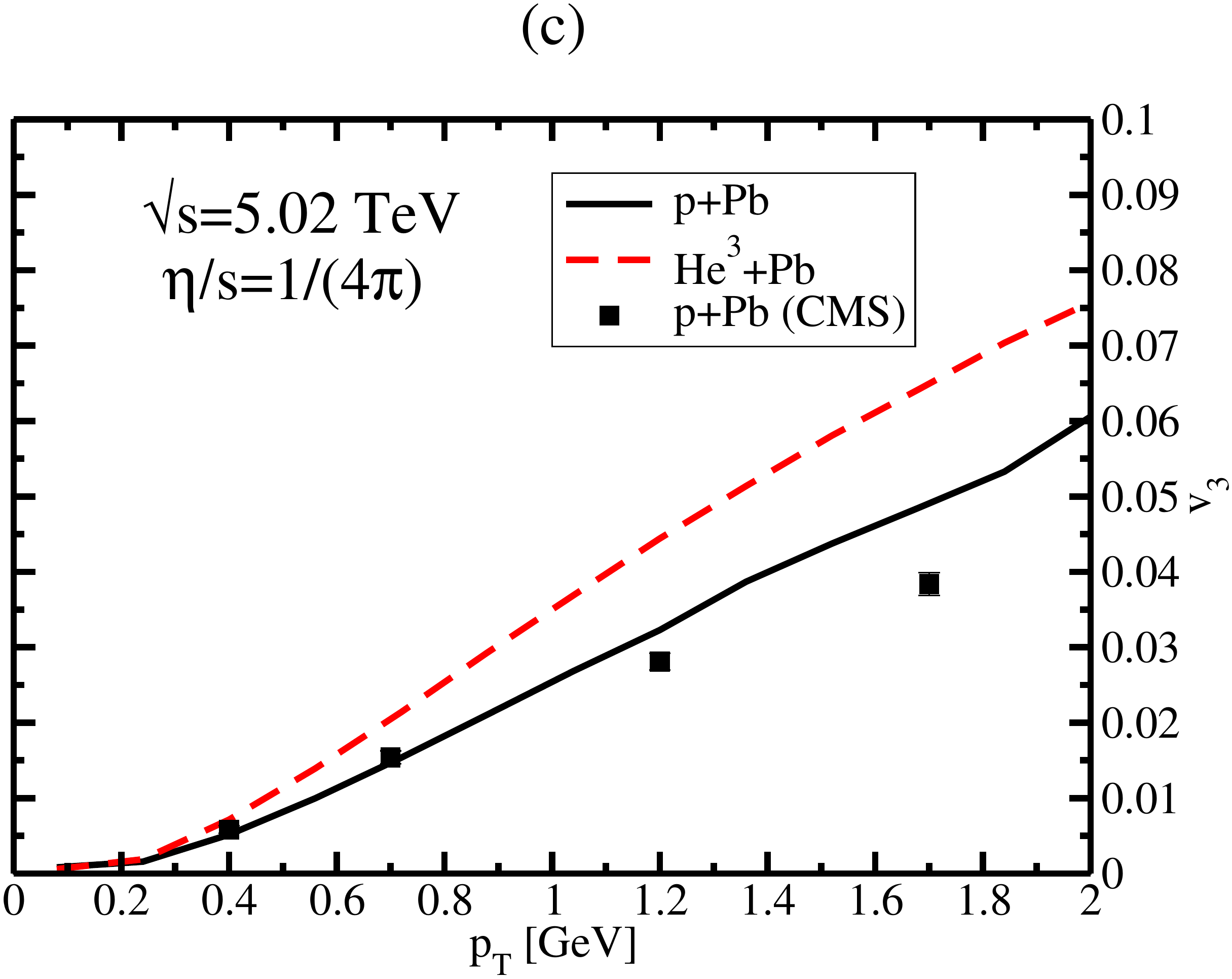}
  \caption{\label{fig:etadep} (Color online) 
(a) pion and proton $v_{2}$ versus $p_{T}$ for \dA and \heA central ($b < 2$ fm) events at RHIC energies using hydrodynamics  for $\eta/s=1/(4\pi)$ in comparison to data for 0-5\% central \dA data from PHENIX \cite{Adare:2014keg}.
(b) pion $v_{3}$ versus $p_{T}$ for \heA central ($b < 2$ fm) events at RHIC energies using viscous hydrodynamics for $\eta/s = 0.2 , 1, 2$ over $4\pi$ as well as result of different smearing parameter $\sigma$.
(c) pion $v_{3}$ versus $p_{T}$ for \pPb and \heA central ($b < 2$ fm) events at LHC energies using viscous hydrodynamics for $\eta/s=1/(4\pi)$ in comparison to data for 0.5-2.5\% central \pPb data from CMS \cite{Chatrchyan:2013nka}.
}\end{figure*}

To reduce the dependence on $T_F$, we have chosen to perform 
a standard Cooper-Frye freezeout at $T=170$ MeV, followed by a hadronic cascade including resonance feed-down corrections~\cite{Novak:2013bqa}.
Figure~\ref{fig:v23results} shows the results for the pion momentum anisotropies $v_{2}$ and $v_{3}$ 
from 400 \pA, 400 \dA, and 400 \heA central ($b < 2$ fm) events run with $\eta/s = 1/4\pi$ and initial 
Gaussian smearing $\sigma = 0.4$ fm and 10,000 cascade events for each of these hydrodynamics runs.  There are substantial event-to-event differences, and the dashed lines indicate the event averaged values.  
The \dA event averaged $v_2$ results are in agreement with the published experimental values~\cite{Adare:2013piz} 
(cf.Figure~\ref{fig:etadep}(a)).
The $v_{2}$ values are larger
in \dA and \heA compared with \pA, and the $v_{3}$ values are largest for \heA as one might expect from the
initial spatial anisotropies. For example, \heA $v_{3}$ at $p_T$=1 GeV exceeds that from \dA and \pA by at least 50 percent.
However, we find that at energies of \sqsn=200 GeV, the system stays within the plasma phase only for 2-3 fm/c. While the effect on this short system lifetime on elliptic flow $v_{2}$ is seemingly rather minor, we find that there is not sufficient time to convert the initial triangularity into flow, resulting in a small overall magnitude of the triangular flow $v_{3}$.

Next we calculate the pion $v_{3}$ as a function of transverse momentum with
viscosity $\eta/s = 0.2/4\pi$, $\eta/s = 1/4\pi$ and $\eta/s = 2/4\pi$.  These results are shown for \heA in Figure~\ref{fig:etadep}(b),
where the increases in viscosity have a dramatic effect in decreasing the $v_{3}$ flow coefficients. 
%
%
%
%
%
%
%
It has been previously observed that an ambiguity exists between a more diffuse initial energy density (thereby
reducing the $\varepsilon_{n}$ values) and a larger viscous damping (thereby reducing the translation of $\varepsilon_{n}$
into $v_{n}$)~\cite{Sorensen:2011hm}.  This issue is significant for the smallest colliding systems, as well as 
ambiguities from sub-nucleonic fluctuations in calculating the initial energy density distribution~\cite{Bzdak:2013zma}.  
For \dA collisions, these differences are highlighted in the $\varepsilon_{n}$ values tabulated with different initial geometry smearing
assumptions in Table I of Ref.~\cite{Adare:2013nff}.  It is notable that the initial condition for starting hydrodynamics at time
$\tau = 0.5$ fm/c depends not only on the initial energy deposition itself, but also any pre-equilibrium dynamics during that first
half fm/c.  

One may posit that the geometric distribution from each
participating nucleon or between participant pairs should be the same in \pA, \dA and \heA at the
same \sqsn.  We thus repeat the above calculation with viscous hydrodynamics $\eta/s = 1/4\pi$ and doubling
the Gaussian smearing to $\sigma = 0.7$ fm.  The change for central events, again defined as impact parameter $b < 2$ fm, 
on the initial $\varepsilon_{2}$ and $\varepsilon_{3}$ mean values is shown in Figure~\ref{fig:moments}.
The results of this calculation on the pion momentum anisotropies  are shown as the orange points in 
Figure~\ref{fig:etadep}(b).  In the case of \heA, comparing $\eta/s=1/(4\pi),\sigma=0.7$ fm and $\eta/s=2/(4\pi),\sigma=0.4$ fm, we find that there is almost complete ambiguity in the case of pion $v_{2}$ and $v_{3}$, but there are strong differences for \pA and \dA (e.g. $v_{2}$ at $p_T$=2 GeV changes by 60 percent for \pA). Thus, the simultaneous measurement
of the flow coefficients in all three colliding systems not only provides key tests of the different explanations of these
phenomena, but also a powerful methodology for discriminating different 
contributions to the final experimental observed anisotropies.
%
%
%
Finally, we point out that increasing \sqsn would result in a longer system lifetime and hence a more pronounced build-up of $v_{3}$. For the case of LHC energies, we find our $v_{3}$ results for \pPb in agreement with published results by CMS and would predict a distinctively higher $v_{3}$ for \hePb collisions at \sqsn=5.02 TeV (cf. Figure~\ref{fig:etadep}(c)).

In summary, we propose a novel set of measurements to control the geometry in small
colliding systems by utilizing \pA, \dA and \heA collisions. 
In particular, the \heA geometry
provides an intrinsic triangularity.
The combination of measurements of different order flow moments in the different geometries will provide
stringent discrimination between effects from the initial state energy deposition and pre-equilibrium dynamics and 
the longer time scale viscous damping during the hydrodynamic phase.

\begin{acknowledgments}
We gratefully acknowledge useful discussions with Shengli Huang, Matt Luzum and Gunther Roland.
We acknowledge funding from the Division of Nuclear Physics of the
U.S. Department of Energy under Grant No. DE-FG02-00ER41152.  
PR acknowledges support from DOE award No. de-sc0008027 and Sloan Award No. BR2012-038. 
MPM acknowledges support from the Los Alamos National Laboratory LDRD project 20120775PRD4. The work of J.L., and J.C. is supported by the U.S. Department of
Energy, Office of Nuclear Physics, and by the NUCLEI SciDAC program.
This research used also resources of the National Energy Research
Scientific Computing Center (NERSC), which is supported by the Office of
Science of the U.S. Department of Energy under Contract No.
DE-AC02-05CH11231.

\end{acknowledgments}

\bibliographystyle{apsrev} \bibliography{smallsystemflow}

\end{document}